\documentclass{epl}
\usepackage{amsmath}
\usepackage{amssymb}
\usepackage{graphicx}

\title{Consistency of nuclear mass formulae}

\author{S. K. Patra, P. Arumugam and L. Satpathy}
\institute{Institute of Physics, Sachivalaya Marg, Bhubaneswar
- 751 005, India.}

 {Binding energies and masses}  {Nuclear
structure models and methods} 

\begin{document}

\maketitle

\begin{abstract}
The general scepticism and loss of faith on the predictive ability
of different mass formulae, arising out of the divergence of their
predictions in unknown regions taken with respect to a reference
mass formula, is successfully dispelled.  When the result 
of relativistic mean field (RMF) theory with a Lagrangian 
common for all nuclei is taken as reference, the divergence 
disappears, and clear trend with strong correlation appears
restoring our faith in general on the predictions of mass formulae,
qualifying them as useful guideline for theoretical and experimental
studies of nuclear phenomena.
\end{abstract}

Mass formulae occupy the centre stage in the research in nuclear 
physics.  The first model of the nucleus is a mass formula
proposed by Bethe and Weiszackar in mid 1930s,  soon after the
discovery of the neutron unravelling the composition of the
nucleus.  It defined some of the key concepts and parameters for 
the first time for the description of nuclear phenomena, and laid
the foundation for their future exploration.  Our
inability to predict the masses of nuclei starting from first
principle, and our present experimental incapability to produce
large majority of them in the laboratory, warrants reliable mass
formulae of nuclei for understanding many phenomena, most notably,
the nucleosynthesis and stellar evolution.  Therefore the
development of nuclear mass formulae has been all along a core
theme of nuclear physics which has implicitly sustained and
nourished the research in diverse areas of nuclear structure and
nuclear reactions.  Over the last 70 years, about a dozen
of mass formulae have been proposed.

\begin{figure}
\centering
\includegraphics[width=0.8\textwidth, clip=true]{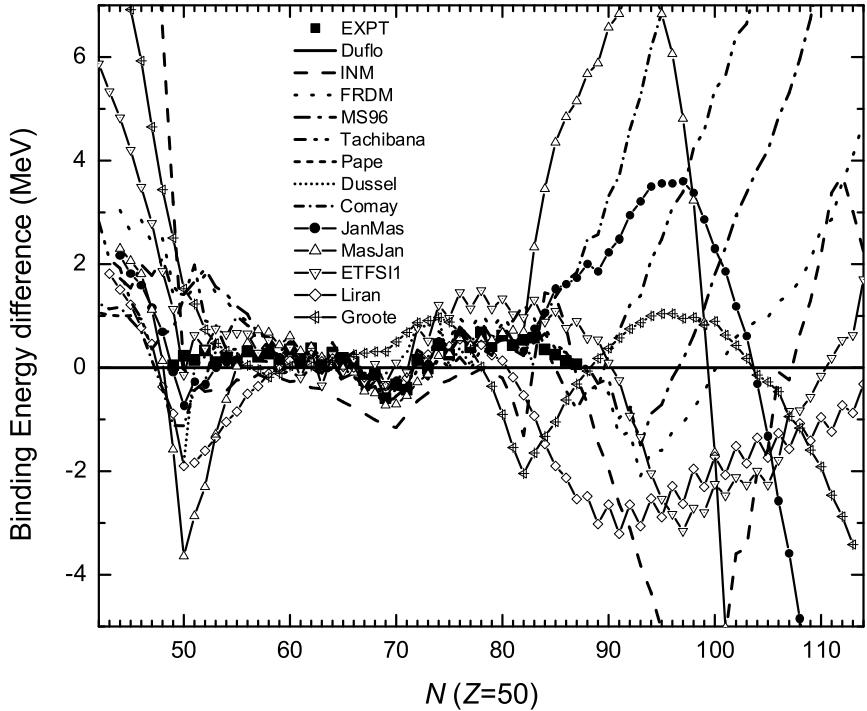}
\caption{Binding energy difference in Sn isotopes with the mass
formula of Duflo and Zuker \cite{Duflo} as reference. 
The data corresponding to Pape, Dussel, MS96, FRDM, Comay, 
Tachibana, JanMas, MasJan are taken from the tables given in
Ref.\ \cite{WebTab}; Liran and Groote are extracted from 
Fig.\ 1 of Ref.\ \cite{Nupecc}; ETFSI1 from Ref.\ 
\cite{ETFSI}.
}
\end{figure}

With the advent of heavy-ion reactions, the prospect of the
exploration of the ``terra incognita'' is very much in the 
realm of possibility.  To the already known about 2000 nuclei,
another 5000 to 7000 nuclei will be added in future by their synthesis in
the laboratory for which unprecedented activity in different
laboratories is underway.  Therefore the predictive ability of
differnt mass formulae is under serious scrutiny.  Figure 1 
represents an often quoted result \cite{Mittig,Nupecc,Sirius} 
on the comparison of the 
predictions of different mass models with experiment on the Tin
isotopes.  Here the difference of the binding energy of various
models and experimental results with respect to the predictions 
of the model of Duflo and Zuker \cite{Duflo}, taken as reference, 
is plotted for the chain of isotopes of Sn with neutron number
varying between 45 and 110.  There is unanimity of all the mass
models on good agreement with experiment in the known region 
close to stability, however, the predictions diverge as one
moves away to unknown regions on either sides where  
measurements have not been feasible.  To see if the same
feature persists in other regions, we have plotted corresponding
diagrams in Fig.\ 2 for the isotopes of $Z=8$, 20, 82, and the
results are similar to Fig.\ 1.
Since the masses of the known regions have been used in the fit
by all the mass models, the agreement with data in these regions is to be 
expected only; what is worrying is in the unknown regions they
do not exhibit a common trend (like rising or falling), but 
diverge without any correlation.  This divergence with the same
intensity is seen when the mass model of M\"oller et al. 
\cite{Moller} is used as reference as seen in Fig.\ 4 of Ref.\ 
\cite{Mittig}.  We have examined the results of microscopic
methods \cite{ETFSI}, to see if they can be used as reference
yielding common trend for the predictions of mass formulae. The
results are only similar to Fig. 1. It is this divergence
which has raised question about the efficacy of the mass models,
and more importantly shaken the faith of the nuclear physics 
community on the predictive ability of the mass models as a
whole.  One is constrained to think whether development of 
different mass models is a parameter game only without
worthwhile theoretical foundation.  In this letter,
we address this issue and show that the above contention is
incorrect, and the mass formulae are in general consistent,
and should be relied upon as guideline.

\begin{figure}
\centering
\includegraphics[width=0.8\textwidth, clip=true]{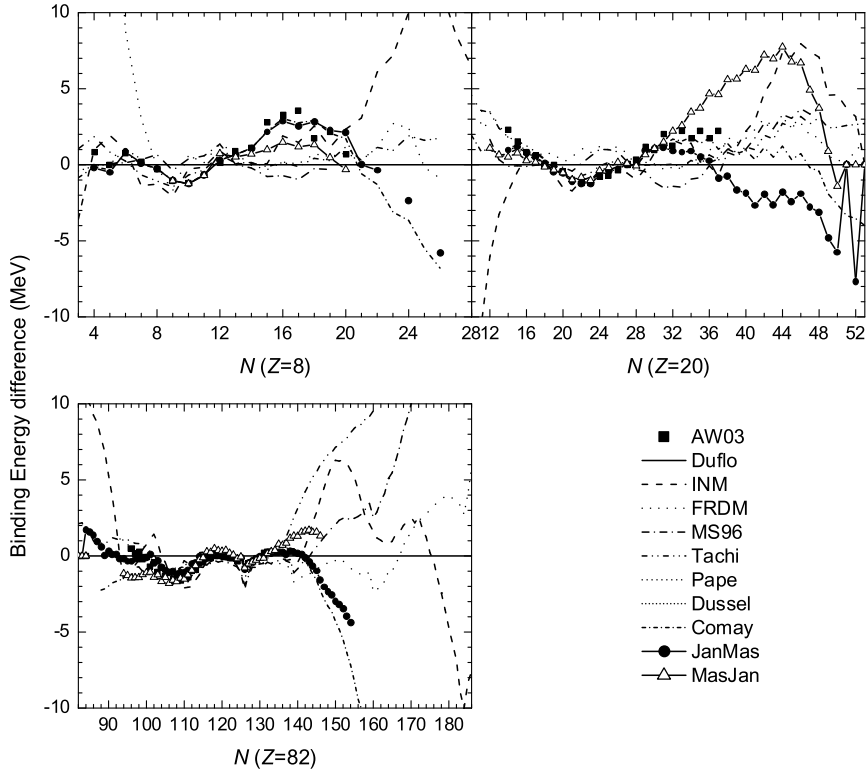}
\caption{Binding energy difference in O, Ca, Sn and Pb isotopes 
with the mass formula of Duflo and Zuker \cite{Duflo} as reference.}
\end{figure}

The basic question that perturbs us is, why the predictions of 
different mass formulae taken with respect to that of Duflo-Zuker 
mass formula as reference, shows the divergence in the unknown 
region ? It is expected that the prediction of each mass formula 
in the unknown region will not agree 100$\%$ with the
future data (when measured) and the discrepancy will go on 
increasing as one moves farther away from the known region. 
Further, it is a common feature with most mass formulae, that the degree of
success varies from region to region even in known domains, and quite
likely, to be more so in unknown domains. This is due to inadequate
accounting of the variation of the characteristic local structure effects
originating from shell, deformation etc. The mass formulae of Duflo-Zuker
and M\"oller et al
are not free from this discomfiture. It is quite likely, 
the degree of accuracy of the predictions of a test mass formula 
region-wise, may not match with those of the reference mass formula,
leading to randomness when the differences are calculated. Therefore 
predictions of different mass formulae taken with respect to reference 
mass formula may show randomness manifesting no common trends 
which appear as divergence in Fig.\ 1. It is like making measurements in
a unsteady co-ordinate system. A fixed common substratum as reference is 
needed with respect to which the predictions of different mass formulae
have to be obtained and their variations be studied. If they do not show
any common trend, and exhibit divergence in the unknown regions,
then one should not repose any faith in their predictions.

It must be recognised that each mass formula, in its own way, tries to 
simulate the effect of the nuclear many-body Hamiltonian in a semi 
phenomenological manner using parameters. All mass models based on 
liquid drop picture assume that the energy of a classical liquid drop added
to the shell correction energy --- using 
a mean field supposed to arise from the
nuclear Hamiltonian --- would represent the true ground-state energy
of a nucleus. The Comay-Kelson \cite{Comay} mass model rests on the
goodness of nuclear mean field picture and the Hartree-Fock (HF) description
of the ground state of the nucleus.
The Duflo-Zuker mass model goes beyond HF assuming the monopole part
of the interaction determines the mean field and the multipole terms 
act as the residual interaction giving rise to configuration
mixing. Thus every mass formula has a foundation of its own rooted in 
the nuclear many-body Hamiltonian, and therefore expected to show, in general a
common trend, although quantitatively they may differ from one another in
their predictions, depending upon the efficacy of their simulation of
nuclear dynamics. With this surmise, we believe, if a common substratum  
with strong theoretical validity for all regions is found in the
form of reference mass model, then the predictions with respect to it
in the unknown region will quite likely show a common trend, but no
divergence.

\begin{figure}
\centering
\includegraphics[width=0.9\textwidth, clip=true]{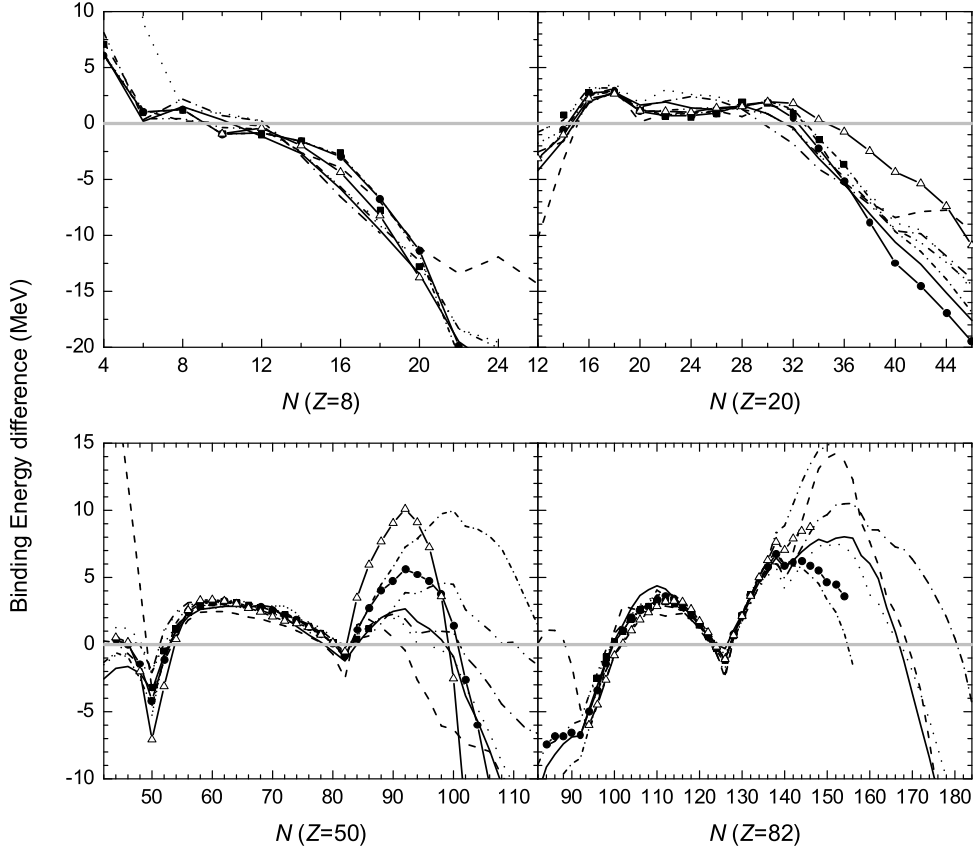}
\caption{Binding energy difference in O, Ca, Sn and Pb isotopes 
with the results of RMF calculations\cite{patra1} [with NL3 
parameter set \cite{NL3}] as reference.}
\end{figure}

The binding energy calculated in RMF theory with an appropriate
field theoretic Lagrangian,
can serve as a common substratum.  With a microscopic
Lagrangian with the parameter set like NL3 \cite{NL3} 
supposed to be valid for all nuclei, and the well defined RMF
formalism being applied without any variation from one region to another,
the binding energies so obtained can qualify to be a good reference.
In Fig.\ 3, the results of such calculations for 10 mass formulae
are presented for the isotopes of $Z=8$, 20, 50 and 82 along with the
experimental data.  It can be seen that for each chain of isotopes
for the O, Ca, Sn and Pb elements, the predictions in the unknown regions, 
show in general a common trend for all the 10 mass formulae although 
quantitatively they differ from one another.  The divergence seen in
Fig.\ 1 and Fig. 2  has disappeared in Fig. 3 
in conformity with our expectation. Similar
results are also obtained with NL-SH, G1 and G2 parameters 
\cite{patra2}. Therefore
the predictions of the mass formulae in the unknown regions 
are not arbitrary or random supposed so far,  rather show definite trend 
which can be used as useful guideline for theoretical and experimental studies.

\begin{figure}
\centering
\includegraphics[width=0.9\textwidth, clip=true]{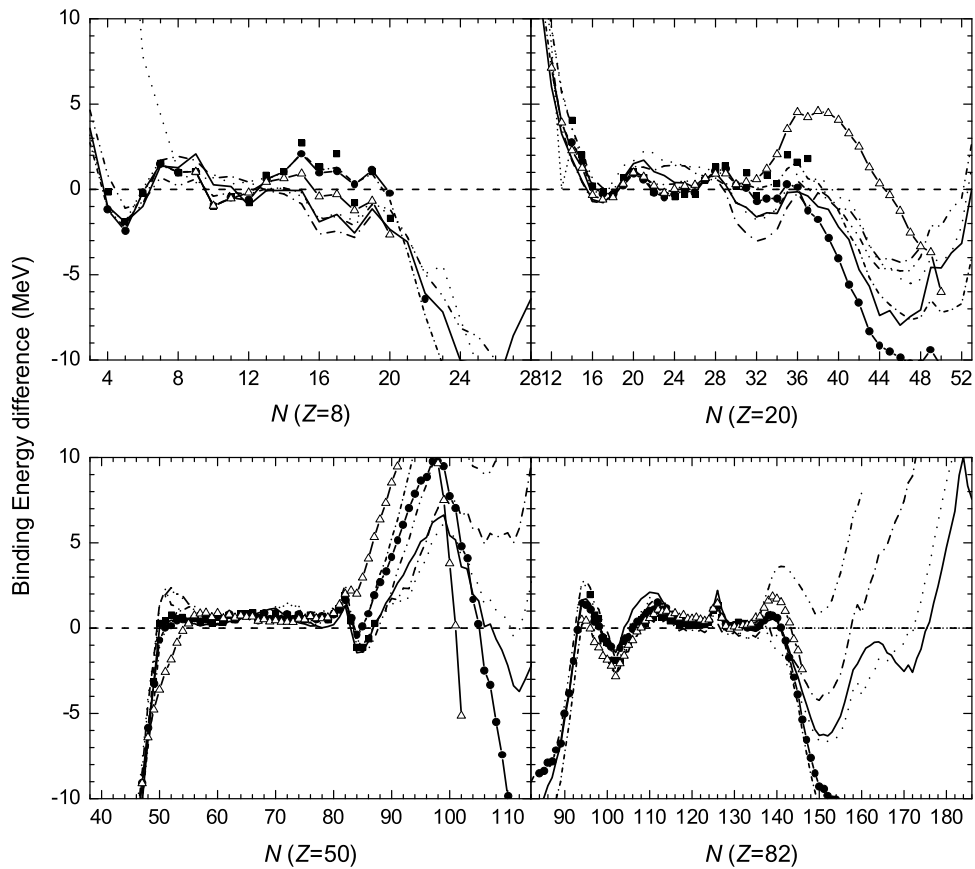}
\caption{Binding energy difference in O, Ca, Sn and Pb isotopes 
with the results of INM mass formula \cite{Satpathy} as reference.}
\end{figure}

The infinite nuclear matter (INM) mass model \cite{Satpathy} 
is based on many-body theoretic foundation.  It
is built in terms of quantum mechanical nuclear liquid rather than
the classical liquid used in Bethe-Weiszacker-like mass models.
The quantum effect is taken into account in this model 
by the use of generalized
Hugenholtz-Van Hove (HVH) theorem \cite{HVH} which also accounts the effect
of three-body force \cite{Satpathy1}. 
In view of its unique success like prediction of
saturation properties and incompressibility \cite{Satpathy1,Satpathy2}
of INM from nuclear
masses, shell quenching \cite{Satpathy3} in large neutron shells N=82, 126
and broadening of stability peninsula 
\cite{Satpathy4}, we use it as the reference mass
formula to see its general validity. The results so obtained
for the 10 mass formulae are presented in Fig.\ 4.  It is 
pleasing to find that no divergences appear in unknown regions,
but common trends --- as found in Fig.\ 3 pertaining to the case
with microscopic RMF calculations as reference --- are seen.  This is 
indicative of the general validity and uniqueness of INM mass formula.

In summary, the notion propagated in the last several years
that the predications of different mass formulae do not show any
correlation and diverge as one moves to the unknown region is
not correct.  This happens because of uneven degree of inaccuracy
in various domains in the predictions of reference mass formula
with respect to which
the results of other mass formulae are calculated.  When predictions of the RMF theory 
with a fixed Lagrangian of general validity for all nuclei is taken, the
divergence disappears and common trends for all mass models
manifest. The INM mass model as reference also reproduces the common
trend.  The general scepticism about the predictability of mass
formulae is then dispelled and our faith on the soundness of
their results is restored.  Their predictions can be used as useful guide
in the theoretical and experimental exploration of the nuclear
phenomena.

\bigskip

We would like to thank Dr. R.K. Choudhury for valuable discussions.


\begin{thebibliography}{99}
\bibitem{Mittig} W. Mittig, P.Roussel-Chomaz and A.C.C.Villari,
Europhysics News (2004) Vol. 35 No. 4 
(http://www.europhysicsnews.com/full/28/article3/article3.html)
\bibitem{Nupecc} NuPECC Report "Nuclear Physics in Europe: Highlights 
and Opportunities" (1997)
$(http://www.nupecc.org/nupecc/report97/report97_final/node6.html)$.
\bibitem{Sirius} SIRIUS Science, CLRC ISBN 0-90376-75-6, December 1998 
(http://marie.surrey.ac.uk/npg/sirius/SIRIUSver3.pdf).
\bibitem{Duflo} J. Duflo and A. Zucker, Nucl. Phys. {\bf A576} (1994) 29;
{\it ibid} Phys. Rev. {\bf C52} (1996) R23.
\bibitem{WebTab} P.E. Haustein, 
                At. Data Nucl. Data Tables {\bf 39} (1988) 185
                (http://ie.lbl.gov/toimass.html).
\bibitem{ETFSI} 
M. Samyn, S. Goriely, P.-H. Heenen, J.M. Pearson and 
F. Tondeur, Nucl. Phys. {\bf A700} (2001) 142;
                S. Goriely, M. Samyn, P.-H. Heenen, J.M. Pearson and 
		F. Tondeur, Phys. Rev. {\bf C66} (2002) 024326;
	        M. Samyn, S. Goriely, and J.M. Pearson, Nucl. Phys.
		{\bf A725} (2003) 69;
                S. Goriely, M. Samyn, M. Bender and J.M. Pearson,
		Phys. Rev. {\bf C68} (2003) 054325;
                M. Samyn, S. Goriely, M. Bender and J.M. Pearson,
                Phys. Rev. {\bf C70} (2004) 044309
                (http://www-astro.ulb.ac.be/Html/masses.html).
\bibitem{Moller} P. M\"oller, J.R. Nix, W.D. Myers and W. Swiatecki,
                 At. Data Nucl. Data Tables {\bf 59} (1995) 185.
\bibitem{Comay}  E. Comay, I. Kelson and A. Zidon, At. Data Nucl. 
                 Data Tables {\bf 39} (1988) 235.
\bibitem{patra1} S.K. Patra and C.R. Praharaj, Phys. Rev.
                 {\bf C44} (1991) 2552.
\bibitem{NL3}    G.A. Lalazissis, J. K\"onig, and P. Ring, Phys.\
                 Rev. \textbf{C55} (1997) 540.
\bibitem{patra2} M. Del Estal, M. Centelles,
                 X. Vi\~nas and S.K. Patra, 
		 Phys. Rev {\bf C63} (2001) 024314.
\bibitem{Satpathy}
	         L. Satpathy, J. Phys. {\bf G13} (1987) 761;
                 R.C. Nayak and L. Satpathy,
	         At. Data Nucl.  Data Tables {\bf 73} (1999) 213;
\bibitem{HVH}    N.H. Hugenholtz and W. Van Hove, Physica (Utrecht) {\bf 24}
                (1958) 363; L. Satpathy and R.C. Nayak, Phys. Rev. Lett.
		{\bf 51} (1983) 1243.
\bibitem{Satpathy1} L. Satpathy, V.S. Uma Maheswaru and R.C. Nayak,
	            Phys. Rep. {\bf 319} (1999) 85.
\bibitem{Satpathy2} R.C. Nayak, V.S. Uma Maheswari and L. Satpathy, 
                    Phys. Rev. {\bf C52} (1995) 711.
\bibitem{Satpathy3} L. Satpathy and R.C. Nayak, 
                    J. Phys. {\bf G24} (1998) 1527.
\bibitem{Satpathy4} L. Satpathy and S.K. Patra, 
		    J. Phys. {\bf G30} (2004) 771. 
\end{thebibliography}
\end{document}